\documentclass[a4paper,fleqn,usenatbib]{mnras}

\usepackage[T1]{fontenc}
\usepackage{ae,aecompl}

\usepackage{graphicx}	
\usepackage{amsmath}	
\usepackage{amssymb}	

\title[Molecules in the Extreme Outflow of a pPN]{\center Early Science with the Large Millimetre Telescope: \\Molecules in the Extreme Outflow of a proto-Planetary Nebula}

\author[A.I. G\'omez-Ruiz et al.]{A.I. G\'omez-Ruiz,$^{1}$\thanks{e-mail: aigomez@inaoep.mx} 
L. Guzman-Ramirez,$^{2,3}$, E. O. Serrano$^{4}$, D. Sanchez-Arguelles$^{4}$, \newauthor 
A. Luna$^{4}$, F. P. Schloerb$^{5}$, G. Narayanan$^{5}$, M. S. Yun$^{5}$, R. Sahai$^{6}$, A. A. Zijlstra$^{7}$, \newauthor
M. Chavez-Dagostino$^{4}$, A. Monta\~na$^{1}$, D. H. Hughes$^{4}$, M. Rodr\'iguez$^{4}$ 
\\
$^{1}$ CONACYT--Instituto Nacional de Astrof\'isica, \'Optica y Electr\'onica, Luis E. Erro 1, 72840 Tonantzintla, Puebla, M\'exico\\
$^{2}$ European Southern Observatory, Alonso de C\'ordova 3107, Casilla 19001, Santiago, Chile\\
$^{3}$ Leiden Observatory, Leiden University, Niels Bohrweg 2, 2333 CA Leiden, The Netherlands\\
$^{4}$ Instituto Nacional de Astrof\'isica, \'Optica y Electr\'onica, Luis E. Erro 1, 72840 Tonantzintla, Puebla, M\'exico\\
$^{5}$ Department of Astronomy, University of Massachusetts, Amherst, MA 01003, USA\\
$^{6}$ Jet Propulsion Laboratory, MS 183-900, California Institute of Technology, Pasadena, CA 91109, USA\\
$^{7}$ Jodrell Bank Centre for Astrophysics, School of Physics and Astronomy, University of Manchester, Manchester M13 9PL, UK\\
}

\date{Accepted XXX. Received YYY; in original form ZZZ}

\pubyear{2016}

\begin{document}
\label{firstpage}
\pagerange{\pageref{firstpage}--\pageref{lastpage}}
\maketitle

\begin{abstract}
Extremely high velocity emission likely related to jets is known to occur in some proto-Planetary Nebulae. However, the molecular complexity of this kinematic component is largely unknown. We observed the known extreme outflow from the proto-Planetary Nebula IRAS 16342-3814, a prototype water fountain, in the full frequency range from 73 to 111 GHz with the RSR receiver on the Large Millimetre Telescope. We detected the molecules SiO, HCN, SO, and $^{13}$CO. All molecular transitions, with the exception of the latter are detected for the first time in this source, and all present emission with velocities up to a few hundred km s$^{-1}$. IRAS 16342-3814 is therefore the only source of this kind presenting extreme outflow activity simultaneously in all these molecules, with SO and SiO emission showing the highest velocities found of these species in proto-Planetary Nebulae. To be confirmed is a tentative weak SO component with a FWHM $\sim$ 700 km s$^{-1}$. The extreme outflow gas consists of dense gas (n$_{\rm H_2} >$ 10$^{4.8}$--10$^{5.7}$ cm$^{-3}$), with a mass larger than $\sim$ 0.02--0.15 M$_{\odot}$. The relatively high abundances of SiO and SO may be an indication of an oxygen-rich extreme high velocity gas. 
\end{abstract}


\section{Introduction}

One of the mysteries in planetary nebulae (PNe) is the morphological changes that transform the spherical circumstellar envelopes (CSEs) of asymptotic giant branch (AGB) stars into highly bipolar/multipolar PNe. To understand the mechanisms of such changes, the short transition phase in between should be explored. At some point in the late-AGB stage, a process (or processes) accelerates and imposes bipolarity upon the slow, spherical AGB winds. What produces bipolarity in these objects and at what stage does bipolarity manifest itself are key questions that remain to be answered.

It has been suggested that fast collimated outflows and jets, active during the proto-planetary nebula (pPN) and/or very late AGB phase, are responsible for the drastic change in the mass-loss geometry and dynamics of the system in transition \citep{Sahai98}. Among the outflows observed in pPNe, there is a category that stands out because of its peculiar kinematics. \citet{Sahai15} coined the term ``Extreme-outflow" to define those pPNe with molecular outflows showing line emission in excess of $\sim$100\,km s$^{-1}$, with a few examples identified by these authors and a few others found in the literature fulfilling this definition. Such spectral feature in pPNe outflows may be the equivalent of the so-called extremely high velocity (EHV) emission observed in protostellar outflows, which is thought to be an unambiguous jet signature \citep{Bachiller96}. Indeed, some studies in pPNe outflows support this similarity \citep[see, e.g.,][]{Balick13}. Star formation outflows presenting this peculiar spectral feature have been used as the perfect tool to study the kinematics of the different components of the outflow process, but in particular allowing a more complete study of the jet (EHV) component, otherwise contaminated by the other outflow components, such as the cavity and the bow shocks \citep[see, e.g.,][]{Lefloch15}. The identification of extreme-outflows from pPNe is therefore relevant, since their study have the potential to put constrains to theoretical models that include jets. 

Water fountain pPNe are a particularly interesting subclass of pPNe whose original distinguishing characteristic is the presence of very high-velocity red and blue-shifted H$_2$O and OH maser features. The velocity separations of the water fountains can be as high as 500\,km s$^{-1}$ \citep{Gomez11}. This velocity spread of the masers may suggest a relation between water fountains and the extreme-outflows traced by thermal molecular lines \citep{Yung16}. IRAS16342-3814 is the nearest \citep[$\sim$2 kpc;][]{Sahai99} and best studied water-fountain. Its morphology has been resolved in the optical, near-infrared, and mid-infrared. Radio observations show water masers spread over a wide range of radial velocities ($>$100\,km s$^{-1}$) \citep{Chong15}. CO (2-1) and CO (3-2) observations reveal a massive, high-velocity molecular outflow \citep{He08,Imai09,Imai12}. The CO line profiles exhibit both, a narrow component with an expansion velocity of 40\,km s$^{-1}$, and wide wings with an expansion velocity of 100\,km s$^{-1}$ \citep{Imai12}.

The molecular complexity of pPNe with outflows is well known \citep[e.g.][]{Sanchez97}, however in most of the cases the molecular emission can not be unambiguously related to the jet component. Water fountains, on the other hand, have been poorly covered by molecular line observations. The EHV emission from pPNe outflows has been little explored in other molecular species than CO. Recent studies in star formation outflows with EHV emission have unveiled a velocity-dependent shock chemistry and a peculiar composition of the EHV gas \citep{Tafalla10}. With this background, we started a project to study the molecular composition of pPNe outflows with known EHV emission from CO lines. In this letter we present wide band observations towards the pPN IRAS16342-3814, with the double aim of providing a molecular census in this prototype water fountain and to study the molecular composition of the EHV gas. 

\section{Observations}
Using the Large Millimetre Telescope Alfonso Serrano in its early science phase, we observed the known EHV outflow pPN IRAS16342-3814 with the Redshift Search Receiver (RSR). Observations were performed on March 19th and 24th, 2016, with a sky opacity, $\tau_{\rm 225 GHz}$, ranging from 0.16 to 0.20, and an instrumental T$_{\rm sys}$ from 109 to 116 K. Observations were centred on the coordinates RA (J2000) $=$ 16:37:39.91, DEC (J2000) $=$ -38:20:17.3; with the OFF beam 39$''$ apart. Pointing accuracy was found to be better than 2$''$. The total ON time integration was 1.5 hrs. 

The RSR is an autocorrelator spectrometer that covers the frequency range of 73--111 GHz, at 31 MHz spectral resolution, which correspond to $\sim$ 100 km s$^{-1}$ at 90 GHz \citep{Erickson07}. In its early science phase, the LMT operates with a 32m active surface, which then results in a Half-Power Beam Width (HPBW) $\sim$ 26$''$ at the centre of the band \citep[{RSR data has been reported in a number of papers, for an example see} ][]{Cybulski16}. Autocorrelations, spectra co-adding, calibration, and baseline removal was made with the DREAMPY (Data REduction and Analysis Methods in PYTHON) software. A careful inspection of each scan shows reasonable flat baselines and therefore only a linear fit in DREAMPY was used to remove the baseline to each 5min spectrum. However, the average 90min spectrum showed a subtle low-frequency continuous wave signal along the full RSR band. In deep ($\sigma <1$mK) RSR spectra, the atmosphere fluctuations have a non-negligible contribution to the baseline, which is reflected into a structured noise spanning across each board. We then constructed a template for this structured signal using a third order Savitsky-Golay filter with a window size of 1GHz, which is significantly greater than the width of the lines in our spectrum (see sect. 3). Our approach is slightly different from the technique used in \citet{Cybulski16}, since we calculate and subtract the template for each individual spectrum rather than applying the filter for the final spectrum only. Atmospheric residuals can be very different from scan to scan, therefore our final average spectrum is less affected by artifact signals produced by subtraction effects that may be produced if only the final spectrum is passed through the filter. The filtered spectrum has an average r.m.s noise of 0.29 mK. Note, however, that the noise is not uniform along the band, and some parts are less noisy. To convert antenna temperature units (T$_A^*$) to Jy, we used a conversion factor (Jy/K) of 6.4 for $\nu$ $<$ 92 GHz and 7.6 for $\nu$ $>$ 92 GHz. Finally, for consistency, the line parameters were obtained from a spectrum re-sampled to the worse spectral resolution corresponding to the lowest frequency transition detected (SiO 2--1; i.e. 108\,km s$^{-1}$). 

\section{Results}

\begin{figure}
\centering
\includegraphics[width=8.5cm,angle=0]{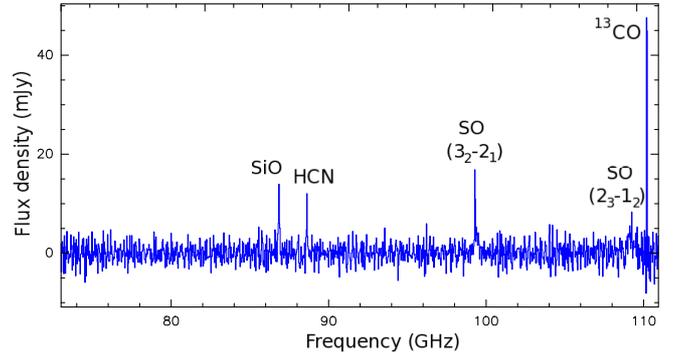}
\caption{LMT/RSR observations of the pPN IRAS 16342-3814. We present the full bandwidth covered by RSR. The species identified are SiO, HCN, SO, and $^{13}$CO.}
\label{rsr}
\end{figure}

\begin{figure} 
\centering
\includegraphics[width=\columnwidth,angle=0, height=11cm]{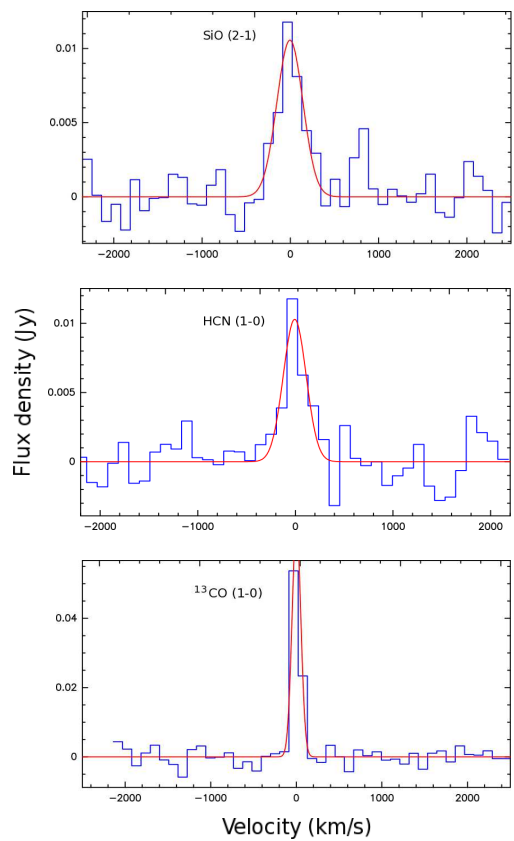}
\caption{Line profiles of the molecules SiO (2--1), HCN (1--0), and $^{13}$CO (1--0), they have all been fitted with a one component Gaussian. Velocity axis is with respect to LSR and the spectral resolution 108 km s$^{-1}$. Using the Gaussian fits we measure the line width, SiO and HCN have a FWHM $>$ 200 km s$^{-1}$, while the $^{13}$CO line has a FWHM of 120 km s$^{-1}$. }
\label{other_lines}
\end{figure}

In Figure \ref{rsr} we present the 3 millimetre spectrum of IRAS 16342-3814. The frequency displayed is topocentric; however, the difference with respect to the rest frame is much smaller than a channel width, and therefore equal to rest frequency within the uncertainties. Table \ref{tab:lines} summarizes the molecular lines detected and their parameters. The following five molecular transitions were detected: SiO (2--1), HCN (1--0), SO (3$_2$--2$_1$) and SO (2$_3$--1$_2$), and $^{13}$CO (1--0). All but $^{13}$CO are first detections in this source. All these lines can be fitted by Gaussian profiles centred close to the systemic velocity (within the uncertainties). Figure \ref{other_lines} shows the SiO (2--1), HCN (1--0), and the $^{13}$CO (1--0) lines with their respective Gaussian fits. In the case of SO (3$_2$--2$_1$), a two-component Gaussian fit seems to be needed to account for the emission (see below). In peak intensity, the strongest line is $^{13}$CO (1--0), while the weakest is SO (2$_3$--1$_2$). SiO (2--1) and HCN (1--0) are approximately similar in peak intensity. Regarding the linewidth, SiO (2--1) is about 40\% wider (Full Width Half Maximum (FWHM) of 322$\pm$47 km s$^{-1}$) than HCN (1--0), while the narrowest lines are SO (2$_3$--1$_2$) and $^{13}$CO (1--0) with FWHM between 108--120 km s$^{-1}$. 

\begin{figure}
\vbox{
\centering
\includegraphics[width=\columnwidth,angle=0, height=3.5cm]{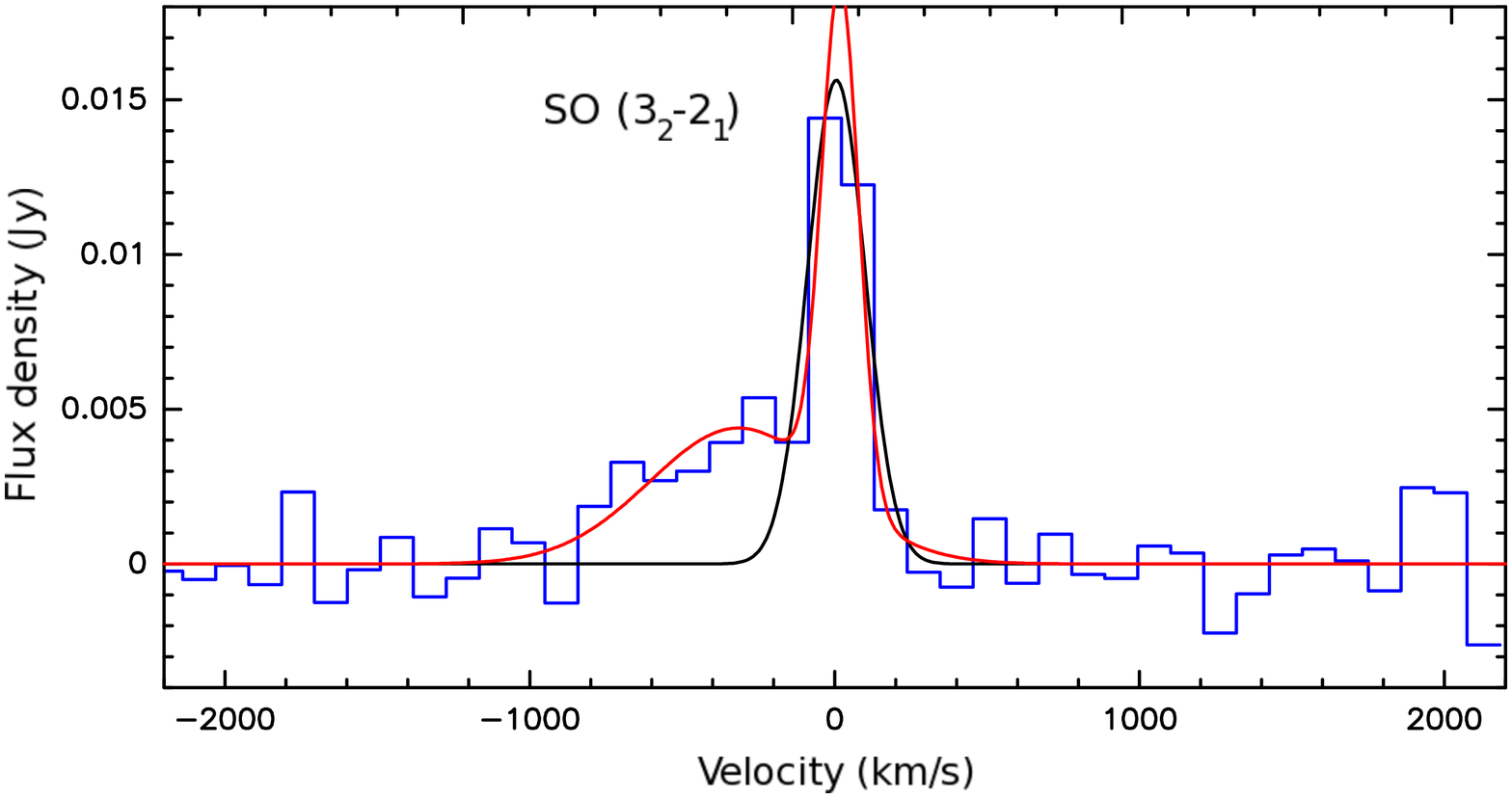}
\includegraphics[width=\columnwidth,angle=0, height=3.5cm]{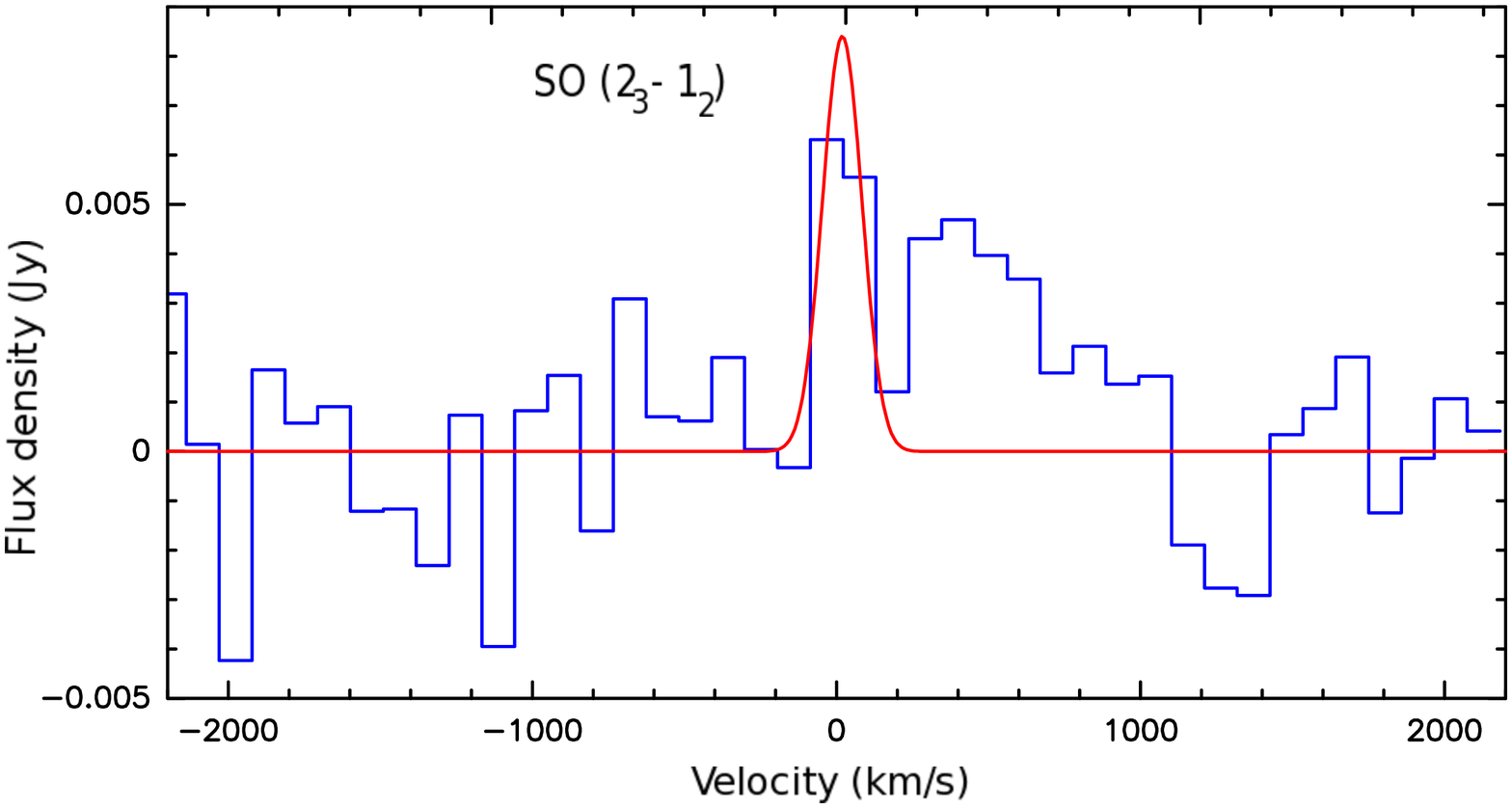} }
\caption{Spectra of the SO lines detected toward the pPN IRAS 16342-3814 and Gaussian fits to their profiles (red line). Velocity axis is with respect to LSR and the spectral resolution 108 km s$^{-1}$. The upper panel shows two different fits, a black one using only one Gaussian fit, and a red fit using two Gaussian components. 
The 2 fits component shows a component in the SO (3$_2$--2$_1$) line with a FWHM of $>$ 600\,km s$^{-1}$, centred at negative velocities, suggesting an extremely fast blue-shifted outflow. For the lower panel we only fitted a narrow component to the SO (2$_3$--1$_2$) line with a FWHM of $\sim$ 108\,km s$^{-1}$.}
\label{SO-2comp}
\end{figure}

In Fig. \ref{SO-2comp} we show Gaussian fits to the SO (3$_2$--2$_1$) and SO (2$_3$--1$_2$) profiles. In a first attempt, we tried to fit a single Gaussian component to the SO (3$_2$--2$_1$) line profile (black line in the upper panel of Fig. \ref{SO-2comp}). The resulting FWHM is 241$\pm$42 km s$^{-1}$. However, we notice a substantial residual coming from the blue-shifted part of the spectrum. Such a residual may suggest the presence of a very fast blue-shifted wing, with a maximum radial velocity (above a 3$\sigma$ level) of $\sim$ $-$700 km s$^{-1}$. A two-component Gaussian fit reduces the residual by a factor of 3 (red line in the upper panel of Fig. \ref{SO-2comp}), resulting in a narrow component at the systemic velocity with a FWHM of 139$\pm$74 km s$^{-1}$ and a wide component with a FWHM of 748$\pm$157 km s$^{-1}$ shifted to negative velocities (peak velocity $-$315$\pm$100 km s$^{-1}$). In the search for other contaminants that may broaden the SO (3$_2$--2$_1$) line, we checked for line transitions inside the range of frequencies that fall into the same range of the line emission (defined within a 3$\sigma$ level). The HCCNC (J$=$10--9) line lies at 99.40 GHz, however this molecule is more common in C-rich objects and IRAS 16342$-$3814, on the other hand, is a O-rich object (as proven by the simultaneous detection of SO and SiO), suggesting such an identification to be unlikely. Furthermore, the frequency of the HCCNC line falls at the high-frequency edge of the broad component. The implications of the wide component is important, since this would represent the highest velocity ever observed in a molecular outflow of a pPN. Although the careful reduction we have performed gives confidence that such a feature is real, given its weakness we prefer to be cautious and defer the confirmation of this feature to future deeper observations (unfortunately no further observations on this object could be taken during the early science season). The SO (2$_3$--1$_2$) line is detected at $\sim$ 4 $\sigma$ in only one channel (note, however, a feature at $\sim$ $+$400 km s$^{-1}$ that we ignore because it barely reaches $\sim$ 3 $\sigma$, and is located in the noisiest part of the band), hence only one Gaussian component is required to fit the emission. This component has a FWHM of 108$\pm$61 km s$^{-1}$ and is centred at the systemic velocity. 

\begin{table*}
\begin{center}
\caption{\em{Molecular lines detected in IRAS 16342$-$3814 with the LMT/RSR and their parameters$^{a}$}}
\begin{tabular}{cccccccc}
\hline \noalign {\smallskip}
Transition & Frequency (GHz) &   HPBW ($''$)  & E$_u$ (K)  &  Line Peak (mJy) & V$_{\rm peak}$ (km s$^{-1}$) & FWHM (km s$^{-1}$)  & $\int$Sdv (mJy km s$^{-1}$)\\
\hline \noalign {\smallskip}
SiO (2--1)        & 86.847  & 27.1 & 6.2   & 10.4  & $-$5(8)& 332(47) & 3677(427)\\
HCN (1--0)        & 88.623  & 26.6 & 12.8  & 10.3  & $-$7(18) & 276(58) & 3023(460)\\
SO (3$_2$--2$_1$) & 99.296  & 23.8 & 9.2   & 15.6  & $+$3(12) & 239(39)  & 3988(453)\\
SO (2$_3$--1$_2$) & 109.252 & 21.6 & 21.1  & 10.3 &  $+$17(11)   & 108(61)  & 1217(395)\\
$^{13}$CO (1--0)  & 110.201 & 21.4 & 5.3   & 66.8 & $+$2(7) & 120(20) & 8572(359)\\
\hline \noalign {\smallskip}
\multicolumn{7}{l}{$a$: Line parameters from Gaussian fits; errors indicated in brackets.}\\
\label{tab:lines}
\end{tabular}
\end{center}
\end{table*}

\section{Discussion}
\subsection{Molecular composition and physical properties}
Before our observations, molecular line data of IRAS 16342$-$3814 revealed the presence of a fast bipolar outflow and a slowly expanding torus/CSE \citep{Imai12}. The CO outflow have a FWHM $\sim$150 km s$^{-1}$ \citep{Imai09}, while the H$_2$O masers show a total velocity spread of 270 km s$^{-1}$, with their proper motions indicating three-dimensional velocities of approximately $\pm$ 180 km s$^{-1}$ \citep{Claussen09}. Our observations show the HCN (1--0) emission with a FWHM similar to the velocity spread of the H$_2$O masers, while the SO emission shows slightly lower values. On the other hand, the SiO emission seems to present higher velocities than the water masers. The $^{13}$ CO (1--0) FWHM is similar the $^{12}$CO measurements.

OH 231.8$+$4.2 is the other pPNe showing high-velocity SO emission with a linewidth $\sim$ 100 km s$^{-1}$ \citep[with a notably asymmetric profile:][]{Claude2000}. Therefore, our detection is the highest velocity SO emission found in a pPN. HCN has been detected at extreme velocities only in CRL 618, with a total velocity range similar to our source \citep[$\sim$250 km s$^{-1}$][]{Sanchez04}. The SiO emission, on the other hand, has been previously detected at extreme velocities of $\leq$ 150 km s$^{-1}$ only in IRAS 08005$-$2356 \citep[albeit in the weak SiO (5--4) and SiO (6--5):][]{Sahai15}. Hence to our knowledge the SiO emission we found in IRAS 16342$-$3814 is the highest velocity reported until now in pPNe. It is important to point out that previous studies show the SiO emission tracing the expanding CSE \citep[see][for a recent study of the SiO emission in evolved stars]{Vicente16}. Our data shows SiO emission with higher velocities, by a factor of three, which must be tracing a different kinematic component, possibly related to the jet, given the similarity with the velocity spread of the water masers. To our knowledge, only IRAS 16342$-$3814 presents EHV emission simultaneously in SiO, SO, HCN and $^{13}$CO.

We assumed LTE conditions in order to calculate the column densities (N) of the molecular species found. In the case of SO, with the two transitions detected, we were able to provide an estimation of the rotational temperature and column density, by employing a rotational diagram analysis \citep[following][]{Claude2000}. In our calculations we have applied beam filling corrections to the line brightness temperatures, assuming a Gaussian source with a size of 0.5$''\times$ 2$''$ \citep[e.g.,][]{Imai12}. We found a T$_{\rm rot}$ in the range of 5-11 K, and N(SO) in the range of 7.0$\times$10$^{16}$ to 4.1$\times$10$^{18}$ cm$^{-2}$. For the other species we also assumed LTE, an excitation temperature of 11 K (i.e. maximum T$_{\rm rot}$ we found for the narrow SO component), and the same source size for beam filling corrections. The results are reported in Table \ref{tab:N_X}.    

Using the SO lines ratio, we constrained the physical conditions of the extreme high velocity gas. We have employed the offline version of RADEX \citep{Tak07} to generate a grid of lines ratios, kinetic temperatures, and H$_2$ volume densities. A more complete description of the RADEX calculations is provided in Appendix A. We assume 2.73 K as the background temperature and a line width of 170 km s$^{-1}$ (average FWHM of the two SO lines). We find that the SO line ratio is not very sensitive to column density within 10$^{11}$ cm$^{-2}$ to $\sim$ 10$^{18}$ cm$^{-2}$. We point out that with only one line ratio it is only possible to provide lower limits to the kinetic temperature and volume density. The observed SO intensity ratio (from Table \ref{tab:lines}: 3988$\pm$453/1217$\pm$395 = 3.3$\pm$1.1) constrains a lower limit to the H$_2$ volume density, n$_{\rm H_2} >$ 10$^{4.8}$--10$^{5.7}$ cm$^{-3}$, and kinetic temperature, T$_{\rm kin} >$ 15--55 K. 

The H$_2$ mass of the SO extreme outflow is determined by using the n$_{\rm H_2}$ lower limit estimated with the LVG analysis from RADEX. \citet{Imai12} modelled the CO emission and determined that the CSE/jet have a size in the range of 6,000--10,000 AU. Assuming that the emission comes from a spheroid with dimensions 6,000 AU $\times$ 6,000 AU $\times$ 10,000 AU, which has a uniform n$_{\rm H_2}$ of 10$^{4.8}$--10$^{5.7}$ cm$^{-3}$ (best value of 10$^{5.3}$ cm$^{-3}$), the mass of such structure should be larger than $\sim$ 0.02--0.15 M$_{\odot}$ (0.06 M$_{\odot}$ for the best value). This mass estimate agrees very well with the mass calculated by \citet[see table 4.]{zijlstra01}, where the dust model of the inner most region gives them a density of 10$^{8}$ cm$^{-3}$, a dust temperature of 80-200 K and a mass of 0.1 M$_{\odot}$. This lower limit to the mass is higher by up to a factor of five than the calculations for a fast outflow as seen from the CO emission \citep[see, e.g.,][]{Sahai06}. Such a dense, very fast and massive material may be related to a jet component. We point out, however, that the SO emission may come from a smaller (shocked) region than the one traced by CO, which then would increase the n$_{\rm H_2}$ estimation. Interferometric mm-wavelength observations with angular resolution below 1$''$ are needed to probe the structure of the SO EHV emission in greater detail.  

In order to test the effect of magnetic fields, we provide a rough estimate of the field intensity needed to broaden the SO line to a few hundred km s$^{-1}$, assuming exclusively the Zeeman effect. Using the formulas given in \citet{Bel89}, we calculated that the minimum field detectable is of the order of mG. For a broadening of $\sim$60-90 MHz the magnetic field intensity needed is $\sim$70G. \cite{vlemings14} estimated $\sim$100G for the upper limit for the magnetic field near the stellar surface for AGB stars, so our estimations are consistent with this limit. However, the resolution of the RSR prevents us from saying anything more conclusive.   

\subsection{Molecular abundances in the EHV gas}

In Table \ref{tab:N_X} we report the molecular abundances (X) computed by using the column density estimations and the expression X=X($^{13}$CO)$\times$N/N($^{13}$CO), where X($^{13}$CO) is assumed to be 2$\times$10$^{-5}$ \citep[following][]{Bujarrabal01}. We notice that the SO abundance (4.0$\times$10$^{-7}$) is consistent with results of the SO wings in OH 231.8$+$4.2 \citet{Claude2000}. The abundance of SiO (5.2$\times$10$^{-7}$) is lower by two orders of magnitude than in IRAS 08005$-$2356, the other known SiO extreme-outflow \citep{Sahai15}. The HCN abundance (9.6$\times$10$^{-8}$) is more similar to the lower limit calculated in CRL 618 for the low-velocity HCN emission ($>$2$\times$10$^{-7}$) than for high-velocity (4$\times$10$^{-6}$) \citep{Sanchez04}. 

\begin{table}
\begin{center}
\caption{\em{Molecular abundances and column densities}}
\begin{tabular}{lcc}
\hline \noalign {\smallskip}
Molecule & N (cm$^{-2}$) &  X$^a$\\
\hline \noalign {\smallskip}
SiO	& 1.4E17	&	5.2E-7\\
HCN	& 2.7E16	&	9.6E-8\\
SO	& 7.0E16	&	4.0E-7\\
$^{13}$CO	& 5.6E18	&	2.0E-5$^b$\\
\hline \noalign {\smallskip}
\multicolumn{3}{l}{$a$: X=X($^{13}$CO)$\times$N/N($^{13}$CO). $b$: Bujarrabal et al. 2001.}\\
\label{tab:N_X}
\end{tabular}
\end{center}
\end{table}

The SiO/HCN abundance ratio is $\sim$ 5, suggesting that C-bearing molecules are depleted in the extreme-outflow. In contrast, the other O-rich pPN with an extreme-outflow, OH 231.8$+$4.2, such abundance ratio is $\sim$ 0.5 for the EHV range \citep[average of $\sim$ 0.3 for the whole line emission range:][]{Sanchez97}. On the other hand, the SiO/SO abundance ratio is close to one, which therefore implicates that the elemental abundances of Si and S are similar; in sharp contrast with OH 231.8$+$4.2 for which such ratio is$\sim$ 10$^{-2}$ \citep{Sanchez97,Claude2000}. A remarkable result is that the highest abundances are those of SiO and SO, which suggests that the extreme high velocity gas is dominated by O-bearing molecules. It is well known that the extreme-outflow in OH 231.8$+$4.2 is more prominent in HCN than in SiO \citep[e.g.,][]{Sanchez97}, which may indicate that this kinematic component, contrary to IRAS 16342$-$3814, is more C-rich than O-rich. In that respect, it is relevant to notice that an O-rich EHV gas in the pPN presented here is more similar to what is found in the star-formation case \citep{Tafalla10}.  

Despite the abundance trend differences noted with respect to OH 231.8$+$4.2, there is still the similarity that both pPNe present a very fast outflow. It has been suggested that the shocks from the extreme-outflow of OH 231.8$+$4.2 may have a major impact on the rich chemistry observed in that source. \citet{Velilla15}, for example, discuss the posibility of molecular reformation after the high-speed shocks, from the interaction of the jet and the CSE, destroy the molecules. They propose, in particular, that additional atoms (such as Si and S) may be released to the gas phase from the dust grains by the actions of the shocks, and when the shocked material has cooled down sufficiently to allow molecular reformation, there is a different proportion of the elements available for the reactions in the post-shocked gas, affecting the abundances of a second-generation of molecules. Such process may also be applicable to the case of IRAS 16342$-$3814. In that regard, we notice that the SiO abundance is close to the range predicted by shock chemistry models \citep[SiO production in the gas phase through the sputtering of Si-bearing material in refractory grain cores:][]{Gusdorf08a}, which may give support to such scenario. However, we point out that the caveat here is that the abundance calculations are dependent on the assumed value of the excitation temperature (that can be different for different molecules), hence a better estimation of such quantity is required before drawing any general conclusion. Also, the limitation of our spectral line data is that it is difficult to separate the contribution from the low-velocity (circumstellar envelope) and high-velocity (jet) components. Hence the abundances reported here are the average of both components, which may not be sufficent for detailed future shock chemistry models. 

\section{Conclusions}

    \begin{enumerate}
      \item Our modest 1.5 hr LMT observations with the RSR probed to be a good strategy to test the chemical content of extreme velocity outflows from pPNe.
      \item With a linewidth of $\sim$ 240 km s$^{-1}$, we found the highest velocity SO outflow in a pPN. 
     \item The HCN emission has a velocity spread similar to CRL618, the other pPN known with high-velocity HCN.
     \item We found the highest velocity SiO emission in a pPN, which may trace a different kinematic component with respect to previous SiO observations in evolved stars, possibly related to a jet.   
\item By using a SO line ratio, we found that the extreme outflow consist of dense gas (n$_{\rm H_2} >$ 10$^{4.8}$--10$^{5.7}$ cm$^{-3}$), with a mass of $\sim$ 0.02--0.15 M$_{\odot}$.
\item We found a weak indication of very fast SO (3$_2$--2$_1$) emission reaching velocities $\sim$ $-$700 km s$^{-1}$, that if confirmed would be the fastest molecular outflow ever found in a proto-Planetary Nebula.
    \end{enumerate}

\section{acknowledgements}

This work would not have been possible without the long-term financial support from the Mexican Science and Technology Funding Agency, Consejo Nacional de Ciencia y Tecnolog\'ia (CONACYT), during the construction and early science phase of the Large Millimetre Telescope Alfonso Serrano, as well as support from the the US National Science Foundation via the University Radio Observatory program, the Instituto Nacional de Astrof\'isica, \'Optica y Electronica (INAOE), and the University of Massachusetts (UMASS). AGR is funded by the program C\'atedras CONACYT para Jovenes Investigadores. LGR is co-funded under the Marie Curie Actions of the European Commission (FP7-COFUND). 

\bibliographystyle{mnras}
\bibliography{biblio}

\newpage
\begin{appendix}

\section{RADEX diagnostic plot}

We used the offline version of RADEX \citep{Tak07} to generate a diagnostic plot of the SO (3$_2$--2$_1$) to SO (2$_3$--1$_2$) intensity ratio, as a function of kinetic temperatures and H$_2$ volume densities. A discussion on the use of these diagnostic plots are presented for different molecules in the appendix of \citet{Tak07}. Linear molecules, such as SO, are density probes at low densities, while at higher densities the line ratios are more sensitive to temperature. However, with no other information than a single line ratio (such as our case), it is only possible to provide lower limits to density and temperature.  

\begin{figure}
\centering
\includegraphics[width=8.5cm,angle=-90]{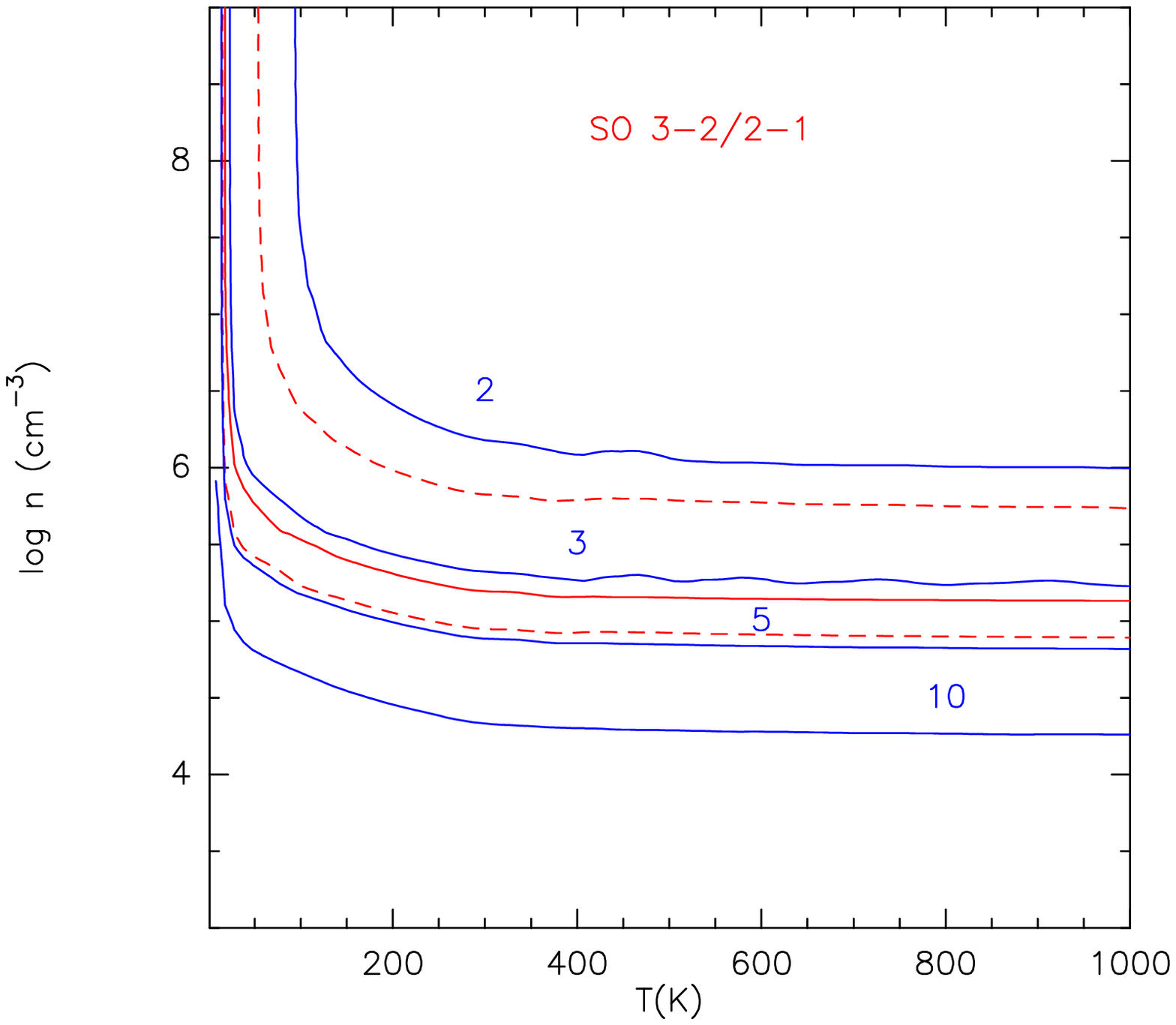}
\caption{Results from the LVG analysis of the SO lines detected. The curves represent the predicted SO (3$_2$--2$_1$) to SO (2$_3$--1$_2$) intensity ratio for a column density of 5$\times$10$^{14}$ cm$^{-2}$. The ratios that match the observed SO line ratio and its errors (3.3$\pm$1.1) are indicated by the solid and dashed red curves, respectively.}
\label{SO-LVG}
\end{figure}

In Fig. \ref{SO-LVG} we show the diagnostic plot for the SO (3$_2$--2$_1$) to SO (2$_3$--1$_2$) intensity ratio, assuming 2.73 K as the background temperature and a line width of 170 km s$^{-1}$ (average FWHM of the two SO lines). The plot shows the results for a column density of 5$\times$10$^{14}$ cm$^{-2}$, however we find that the SO line ratio is not very sensitive to column density within the range 10$^{11}$ cm$^{-2}$ to $\sim$ 10$^{18}$ cm$^{-2}$. The ratios obtained from this run varies from 1.77 to 145.57. The SO intensity ratio that match the observed ratio (from Table \ref{tab:lines}: 3988$\pm$453/1217$\pm$395 = 3.3$\pm$1.1) constrains a lower limit to the H$_2$ volume density, n$_{\rm H_2} >$ 10$^{4.8}$--10$^{5.7}$ cm$^{-3}$, and kinetic temperature, T$_{\rm kin} >$ 15--55 K. 

\end{appendix}

\bsp	
\label{lastpage}
\end{document}